\def\prg#1{\medskip{\bf #1}}          
\def\lra{\leftrightarrow}             
\def\ra{\rightarrow}                  
\def\pd{\partial}                     \def\pgt{{\scriptstyle\rm PGT}}
\def\dis{\displaystyle}               
                 \def\grl{{GR$_\Lambda$}}
\def\Leff{\hbox{$\mit\L_{\hspace{.6pt}\rm eff}$}}
\def\bul{\raise.5ex\hbox{\vrule height.3ex width1ex}\hspace{3pt}}
\def\a{\alpha}        \def\b{\beta}         \def\g{\gamma}
\def\d{\delta}        \def\eps{{\epsilon}}  \def\ve{\varepsilon}
         \def\th{\theta}       
        \def\l{\lambda}       \def\m{\mu}
\def\n{\nu}           \def\om{\omega}       \def\p{\pi}
\def\r{\rho}          \def\s{\sigma}        \def\t{\tau}
\def\vphi{\varphi}
\def\G{{\mit\Gamma}}  \def\D{{\mit\Delta}}  \def\L{{\mit\Lambda}}
\def\Om{{\mit\Omega}}
   \def\tR{{\tilde R}}   \def\tG{{\tilde G}}
\def\tH{{\tilde H}}   \def\hH{{\hat H}}
     \def\cM{{\cal M }}    \def\cO{{\cal O}}
\def\cE{{\cal E}}     \def\cH{{\cal H}}     \def\cK{{\cal K}}
     \def\cB{{\cal B}}
\def\hcO{{\hat\cO}}   \def\hcH{\hat\cH}     \def\bL{{\bar L}}
\def\bcH{\bar\cH}     \def\bcK{\bar\cK}     \def\bD{{\bar D}}
\def\nn{\nonumber}                    \def\vsm{\vspace{-9pt}}
\def\be{\begin{equation}}             \def\ee{\end{equation}}
\def\ba#1{\begin{array}{#1}}          \def\ea{\end{array}}
\def\bea{\begin{eqnarray} }           \def\eea{\end{eqnarray} }
\def\lab#1{\label{eq:#1}}             \def\eq#1{(\ref{eq:#1})}
\def\bsubeq{\begin{subequations}}     \def\esubeq{\end{subequations}}
\def\bitem{\begin{itemize}}           \def\eitem{\end{itemize}}

\documentclass[12pt,twoside]{article}
\usepackage{equations}

\date{}
\begin{document}

\title{Canonical structure of 3D gravity with torsion\footnote{
Invited contribution to appear in {\it Progress in General Relativity
and Quantum Cosmology\/}, vol. 2, ed. Frank Columbus (Nova Science
Publishers, New York, 2005).}}
\author{M. Blagojevi\'c and B. Cvetkovi\'c\footnote{
Email addresses: mb@phy.bg.ac.yu, cbranislav@phy.bg.ac.yu}  \\
{\small Institute of Physics, P. O. Box 57, 11001 Belgrade, Serbia}}

\maketitle

\begin{abstract}
We study the canonical structure of the topological 3D gravity with
torsion, assuming the anti-de Sitter asymptotic conditions. It is shown
that the Poisson bracket algebra of the canonical generators has the
form of two independent Virasoro algebras with classical central
charges. In contrast to the case of general relativity with a
cosmological constant, the values of the central charges are different
from each other.
\end{abstract}

\section{Introduction}
\setcounter{equation}{0}

Faced with enormous difficulties to properly understand fundamental
dynamical properties of gravity, such as the nature of classical
singularities and the problem of quantization, one is naturally led
to consider technically simplified models with the same conceptual
features. An important and useful model of this type is 3D gravity
\cite{1,2}. In the last twenty years, 3D gravity has become an active
research area, with a number of outstanding results. Here, we focus
our attention on a particular line of development, characterized by
the following achievements. In 1986, Brown and Henneaux introduced
the so-called anti-de Sitter (AdS) asymptotic conditions in their
study of 3D general relativity with a cosmological constant (\grl)
\cite{3}. They showed that the related behavior of the gravitational
field allows for an extremely rich asymptotic structure---the
conformal symmetry described by two independent canonical Virasoro
algebras with classical central charges. Soon after that, Witten
rediscovered and further explored the fact that \grl\ in 3D can be
formulated as a Chern-Simons gauge theory \cite{4}. The equivalence
between gravity and an ordinary gauge theory was shown to be crucial
for our understanding of quantum gravity. Then, in 1993, we had the
discovery of the BTZ black hole \cite{5}, with a far-reaching impact
on the development of 3D gravity. All these ideas have had a
significant influence on our understanding of the quantum nature of
3D black holes [2,6-13].

Following a widely spread belief that general relativity is the most
reliable approach for studying the gravitational phenomena, the
analysis of these issues has been carried out mostly in the realm of
{\it Riemannian\/} geometry. However, there is a more general
conception of gravity based on {\it Riemann-Cartan\/} geometry, in
which both the curvature and the torsion characterize the structure
of gravity (see, for instance, Refs. \cite{14,15}). Riemann-Cartan
geometry has been used in the context of 3D gravity since the early
1990s [16-18], with an idea to explore the influence of geometry on
the dynamics of gravity. Recently, new advances in this direction
have been achieved [19-24].

Asymptotic conditions are an intrinsic part of the canonical
formalism, as they define the phase space in which the canonical
dynamics takes place. Their influence on the dynamics is particularly
clear in topological theories, where the propagating degrees of
freedom are absent, and the only non-trivial dynamics is bound to
exist at the asymptotic boundary. General action for topological 3D
gravity with torsion, based on Riemann-Cartan geometry of spacetime,
has been proposed by Mielke and Baekler \cite{16,17}. The objective
of the present paper is to investigate the canonical structure of the
general topological 3D gravity with torsion, including its asymptotic
behavior, in the AdS sector of the theory. This will generalize the
results of Refs. \cite{3,4} and \cite{20}, where the specific choice
of parameters corresponds to Riemannian and telaparallel vacuum
geometry, respectively. Combining this approach with another
interesting result, the existence of the Riemann-Cartan black hole
\cite{19,22}, we shall be able to explore the full influence of
torsion on the canonical and asymptotic structure of 3D gravity.

The paper is  organized as follows. In Sect. 2 we review some basic
features of Riemann--Cartan spacetime as the proper geometric arena
for 3D gravity with torsion, and discuss the field equations derived
from the Mielke-Baekler action. In Sect. 3 we describe the
Riemann-Cartan black hole solution, a generalization of the BTZ black
hole. Then, in Sect. 4, we introduce the concept of asymptotically
AdS configuration, and derive the related asymptotic symmetry, which
turns out to be the same as in general relativity---the conformal
symmetry. In the next section, the asymptotic structure of the theory
is incorporated into the Hamiltonian formalism by calculating the
Poisson bracket (PB) algebra of the canonical generators. It has the
form of two independent Virasoro algebras with classical central
charges, the values of which differ from each other, in contrast to
what we have in Riemannian \grl\ and the teleparallel theory
\cite{3,20}. Finally, Sect. 7 is devoted to concluding remarks, while
Appendices contain some technical details.

Our conventions are given by the following rules: the Latin
indices refer to the local Lorentz frame, the Greek indices refer
to the coordinate frame; the first letters of both alphabets
$(a,b,c,...;$ $\a,\b,\g,...)$ run over 1,2, the middle alphabet
letters $(i,j,k,...;\m,\n,\l,...)$ run over 0,1,2; the signature
of spacetime is $\eta=(+,-,-)$; totally antisymmetric tensor
$\ve^{ijk}$ and the related tensor density $\ve^{\m\n\r}$ are both
normalized so that $\ve^{012}=1$.

\section{Topological 3D gravity with torsion}
\setcounter{equation}{0}

Theory of gravity with torsion can be formulated as Poincar\'e gauge
theory (PGT), with an underlying spacetime structure described by
Riemann-Cartan geometry \cite{14,15}.

\prg{PGT in brief.} The basic gravitational variables in PGT are the
triad field $b^i$ and the Lorentz connection $A^{ij}=-A^{ji}$
(1-forms). The field strengths corresponding to the gauge potentials
$b^i$ and $A^{ij}$ are the torsion $T^i$ and the curvature $R^{ij}$
(2-forms): $T^i= db^i+A^i{_m}\wedge b^m$,
$R^{ij}=dA^{ij}+A^i{_m}\wedge A^{mj}$. Gauge symmetries of the theory
are local translations and local Lorentz rotations, parametrized by
$\xi^\m$ and $\ve^{ij}$.

In 3D, we can simplify the notation by introducing the duals of
$A^{ij}$, $R^{ij}$ and $\ve^{ij}$:
$$
\om_i=-\frac{1}{2}\,\ve_{ijk}A^{jk}\, ,\qquad
  R_i=-\frac{1}{2}\,\ve_{ijk}R^{jk}\, ,\qquad
\th_i=-\frac{1}{2}\,\ve_{ijk}\ve^{jk}\, .
$$
In local coordinates $x^\m$, we can expand the triad and the
connection 1-forms as $b^i=b^i{_\m}dx^\m$, $\om^i=\om^i{}_\m dx^\m$.
Gauge transformation laws have the form
\bea
\d_0 b^i{_\m}&=& -\ve^i{}_{jk}b^j{}_{\m}\th^k-(\pd_\m\xi^\r)b^i{_\r}
     -\xi^\r\pd_\r b^i{}_\m\equiv\d_\pgt b^i{}_\m\, ,      \nn\\
\d_0\om^i{_\m}&=& -(\pd_\m\th^i+\ve^i{}_{jk}\om^j{_\m}\th^k)
     -(\pd_\m\xi^\r)\om^i{_\r}
     -\xi^\r\pd_\r\om^i{}_\m\equiv\d_\pgt\om^i{}_\m\, ,    \lab{2.1}
\eea
and the field strengths are given as
\bea
&&T^i=\nabla b^i\equiv db^i+\ve^i{}_{jk}\om^j\wedge b^k
     =\frac{1}{2}T^i{}_{\m\n}dx^\m\wedge dx^\n\, ,         \nn\\
&&R^i=d\om^i+\frac{1}{2}\,\ve^i{}_{jk}\om^j\wedge\om^k
      =\frac{1}{2}R^i{}_{\m\n}dx^\m\wedge dx^\n\, ,        \lab{2.2}
\eea
where $\nabla=dx^\m\nabla_\m$ is the covariant derivative.

To clarify the geometric meaning of the above structure, we introduce
the metric tensor as a specific, bilinear combination of the triad
fields:
\bea
&&g=\eta_{ij}b^i\otimes b^j=g_{\m\n}dx^\m\otimes dx^\n \, ,\nn\\
&& g_{\m\n}=\eta_{ij}b^i{_\m}b^j{_\n}\, ,
            \qquad \eta_{ij}=(+,-,-)\, .                   \nn
\eea
Although metric and connection are in general independent
dynamical/geometric variables, the antisymmetry of $A^{ij}$ in PGT is
equivalent to the so-called {\it metricity condition\/}, $\nabla
g=0$. The geometry whose connection is restricted by the metricity
condition (metric-compatible connection) is called {\it
Riemann-Cartan geometry\/}. Thus, PGT has the geometric structure of
Riemann-Cartan space.

The connection $A^{ij}$ determines the parallel transport in the local
Lorentz basis. Being a true geometric operation, parallel transport is
independent of the basis. This property is incorporated into PGT via
the so-called {\it vielbein postulate\/}, which implies the identity
\be
A_{ijk}=\D_{ijk}+K_{ijk}\, ,                               \lab{2.3}
\ee
where $\D$ is Riemannian (Levi-Civita) connection, and
$K_{ijk}=-\frac{1}{2}(T_{ijk}-T_{kij}+T_{jki})$ is the contortion.

\prg{Topological action.} In general, gravitational dynamics is
defined by Lagrangians which are at most quadratic in field strengths.
Omitting the quadratic terms, Mielke and Baekler proposed a {\it
topological\/} model for 3D gravity \cite{16,17}, with an action of the
form
\bsubeq\lab{2.4}
\be
I=aI_1+\L I_2+\a_3I_3+\a_4I_4+I_M\, ,                      \lab{2.4a}
\ee
where $I_M$ is a matter contribution, and
\bea
&&I_1=2\int b^i\wedge R_i\, ,                              \nn\\
&&I_2=-\frac{1}{3}\,\int\ve_{ijk}b^i\wedge b^j\wedge b^k\,,\nn\\
&&I_3=\int\left(\om^i\wedge d\om_i
  +\frac{1}{3}\ve_{ijk}\om^i\wedge\om^j\wedge\om^k\right)\,,\nn\\
&&I_4=\int b^i\wedge T_i\, .                               \lab{2.4b}
\eea
\esubeq
The first term, with $a=1/16\pi G$, is the usual Einstein-Cartan
action, the second term is a cosmological term, $I_3$ is the
Chern-Simons action for the Lorentz connection, and $I_4$ is an
action of the translational Chern-Simons type. The Mielke-Baekler
model can be thought of as a natural generalization of Riemannian
\grl\ (with $\a_3=\a_4=0$) to a topological gravity theory in
Riemann-Cartan spacetime.

\prg{Field equations.} Variation of the action with respect to triad
and connection yields the gravitational field equations:
\bea
&&\ve^{\m\n\r}\left[aR_{i\n\r}+\a_4T_{i\n\r}
  -\L\ve_{ijk}b^j{_\n}b^k{_\r}\right]=\t^\m{_i} \, ,       \nn\\
&&\ve^{\m\n\r}\left[\a_3R_{i\n\r}+ aT_{i\n\r}
  +\a_4\ve_{ijk}b^j{_\n}b^k{_\r}\right]=\s^\m{_i}\, ,      \nn
\eea
where $\t^\m{}_i=-\d I_M/\d b^i{}_\m$ and
$\s^\m{}_i=-\d I_M/\d\om^i{}_\m$ are the matter energy-momentum and
spin currents, respectively. For our purposes---to study the
canonical structure of the theory in the asymptotic region---it is
sufficient to consider the field equations in vacuum, where
$\t=\s=0$.  In the sector $\a_3\a_4-a^2\ne 0$, these equations take
the simple form
\bsubeq\lab{2.5}
\bea
T_{ijk}=p\ve_{ijk}\, ,                                     \lab{2.5a}\\
R_{ijk}=q\ve_{ijk}\, ,                                     \lab{2.5b}
\eea
\esubeq
where
$$
p=\frac{\a_3\L+\a_4 a}{\a_3\a_4-a^2}\, ,\qquad
q=-\frac{(\a_4)^2+a\L}{\a_3\a_4-a^2}\, .
$$
Thus, the vacuum configuration is characterized by constant torsion
and constant curvature.

In Riemann-Cartan spacetime, one can use the identity \eq{2.3} to
express the curvature $R^{ij}{}_{\m\n}(A)$ in terms of its Riemannian
piece $\tR^{ij}{}_{\m\n}\equiv R^{ij}{}_{\m\n}(\D)$ and the
contortion:
$$
R^{ij}{}_{\m\n}(A)=\tR^{ij}{}_{\m\n}+\left[
  \nabla_\m K^{ij}{}_\n-K^i{}_{m\m}K^{mj}{_\n}-(\m\lra\n)\right]\, .
$$
This relation, combined with the field equations \eq{2.5}, leads to
\be
\tR^{ij}{} _{\m\n}= -\Leff(b^i{_\m} b^j{_\n}-b^i{_\n} b^j{_\m}),
\qquad \Leff\equiv q-\frac{1}{4}p^2\, ,                    \lab{2.6}
\ee
where $\Leff$ is the effective cosmological constant. Equation \eq{2.6}
can be considered as an equivalent of the second field equation \eq{2.5b}.
Looking at \eq{2.6} as an equation for the metric, one concludes that
our spacetime has maximally symmetric metric \cite{25}:
\bitem
\item[\bul] for $\Leff<0$ ($\Leff>0$), the spacetime manifold is
anti-de Sitter (de Sitter).
\eitem

There are two interesting special cases of the general Mielke-Baekler
model, which have been studied in the past.
\bitem
\item[\bul] For $\a_3=\a_4=0$, the vacuum geometry becomes
{\it Riemannian\/}, $T_{ijk}=0$. This choice corresponds to \grl\
\cite{3,4}; \vsm
\item[\bul] for $(\a_4)^2+a\L=0$, the vacuum geometry is
{\it teleparallel\/}, $R_{ijk}=0$.  The vacuum field equations are
``geometrically dual" to those of \grl\ \cite{20}.
\eitem
In the present paper, we shall investigate the general model \eq{2.4}
with $\a_3\a_4-a^2\ne 0$, assuming that the effective cosmological
constant is negative (anti-de Sitter sector):
\be
\Leff\equiv-\frac{1}{\ell^2}<0 \, .                        \lab{2.7}
\ee
The de Sitter sector with $\Leff>0$ is left for the future studies.

\section{Exact vacuum solutions}
\setcounter{equation}{0}

Some aspects of the canonical analysis rely on the existence of
suitable asymptotic conditions. A proper choice of these conditions
is based, to some extent, on the knowledge of exact classical
solutions in vacuum. For the Mielke-Baekler model \eq{2.4}, these
solutions are well known \cite{19,22}. Their construction can be
described by the following set of rules:
\bitem
\item[\bul] For a given $\Leff$, use Eq. \eq{2.6} to find a
solution for the metric. This step is very simple, since the metric
structure of maximally symmetric spaces is well known \cite{25}. \vsm
\item[\bul] Given the metric, find a solution for the triad field,
such that $g=\eta_{ij}\,b^i\otimes b^j$. \vsm
\item[\bul] Finally, use Eq. \eq{2.5a} to determine the connection
$\om^i$.
\eitem
For exact solutions with non-vanishing sources, the reader can consult
Ref. \cite{24}.

\prg{Riemann-Cartan black hole.} For $\Leff<0$, equation \eq{2.6} has
a well known solution for the metric --- the BTZ black hole \cite{5}.
Using the static coordinates $x^\m=(t,r,\vphi)$ (with
$0\le\vphi<2\pi$), and units $4G=1$, it is given as
\bea
&&ds^2=N^2dt^2-N^{-2}dr^2-r^2(d\vphi+N_\vphi dt)^2\, ,     \nn\\
&&N^2=\left(-2m+\frac{r^2}{\ell^2}+\frac{J^2}{r^2}\right)\, ,
  \qquad N_\vphi=\frac{J}{r^2}\, .                         \lab{3.1}
\eea
The parameters $m$ and $J$ are related to the conserved
charges---energy and angular momentum. Since the triad field
corresponding to \eq{3.1} is determined only up to a local Lorentz
transformation, we can choose $b^i$ to have the simple, ``diagonal"
form:
\bsubeq\lab{3.2}
\be
b^0=Ndt\, ,\qquad b^1=N^{-1}dr\, ,\qquad
b^2=r\left(d\vphi+N_\vphi dt\right)\, .                    \lab{3.2a}
\ee
Then, the connection is obtained by solving the first field equation
\eq{2.5a}:
\bea
&&\om^0=N\left(\frac{p}{2}dt-d\vphi\right)\, ,\qquad
  \om^1=N^{-1}\left(\frac{p}{2}+\frac{J}{r^2}\right)dr\, , \nn\\
&&\om^2= -\left(\frac{r}{\ell}
  -\frac{p\ell}{2}\frac{J}{r}\right)\frac{dt}{\ell}
  +\left( \frac{p}{2}r-\frac{J}{r}\right)d\vphi \, .       \lab{3.2b}
\eea
\esubeq
Equations \eq{3.2} define the {\it Riemann-Cartan\/} black hole.

\prg{Riemann-Cartan AdS solution.} In Riemannian geometry with
negative $\L$, the ge\-ne\-ral solution with maximal number of
Killing vectors is called the AdS solution \cite{5,25}.  Although AdS
solution and the black hole are locally isometric, they are globally
distinct. The AdS solution can be obtained from \eq{3.1} by the
replacement $J=0$, $2m=-1$.

Similarly, there is a general solution with maximal symmetry in
Riemann-Cartan ge\-o\-me\-try, the {\it Riemann-Cartan\/} AdS solution.
It can be obtained from the black hole \eq{3.2} by the same replacement
($J=0$, $2m=-1$). Using the notation $f^2\equiv 1+r^2/\ell^2$, we have:
\bsubeq\lab{3.3}
\bea
&&b^0=fdt\, ,\hspace{82pt} b^1=f^{-1}dr\, ,
             \qquad  b^2=rd\vphi\, ,                       \lab{3.3a}\\
&&\om^0=f\left(\frac{p}{2}dt-d\vphi\right)\, ,\qquad
\om^1=\frac{p}{2f}dr\, , \qquad
\om^2= -\frac{r}{\ell}
\left(\frac{dt}{\ell}-\frac{p\ell}{2}d\vphi\right)\, .     \lab{3.3b}
\eea
\esubeq

In order to understand symmetry properties of \eq{3.3}, we note that
the form-invariance of a given field configuration in Riemann-Cartan
geometry is defined by the requirements $\d_0 b^i{_\m}=0$,
$\d_0\om^i{_\m}=0$, which differ from the Killing equation in
Riemannian geometry, $\d_0 g_{\m\n}=0$ ($\d_0$ is the PGT analogue of
the geometric Lie derivative). When applied to the Riemann-Cartan AdS
solution \eq{3.3}, these requirements restrict ($\xi^\m,\th^i$) to
the subspace defined by the basis of six pairs
($\xi^\m_{(k)},\th^i_{(k)}$) ($k=1,\dots,6$), given in Appendix A.
The related symmetry group is the six-dimensional AdS group
$SO(2,2)$.

\section{Asymptotic conditions}
\setcounter{equation}{0}

Spacetime outside localized matter sources is described by the vacuum
solutions of the field equations \eq{2.5}. Thus, matter has no
influence on the local properties of spacetime in the source-free
regions, but it can change its global properties. On the other hand,
global properties of spacetime affect symmetry properties of the
asymptotic configurations, and consequently, they are closely related
to the gravitational conservation laws.

In 3D gravity with $\Leff<0$, maximally symmetric AdS solution \eq{3.3}
has the role ana\-lo\-gous to the role of Minkowski space in the $\Leff=0$
case. Following this analogy, we could choose \eq{3.3} to be the field
configuration to which all the dynamical variables approach in such a
way, that the asymptotic symmetry is $SO(2,2)$, the maximal symmetry of
\eq{3.3}. However, such an assumption would exclude the important black
hole geometries, which are not $SO(2,2)$ invariant. Having an idea to
maximally relax the asymptotic conditions and enlarge the set of
asymptotic states (and the relevant group of symmetries), we introduce
the concept of the {\it AdS asymptotic behavior\/}, based on the
following requirements \cite{3,26}:
\bitem
\item[(a)] asymptotic configurations should include the black hole
           geometries; \vsm
\item[(b)] they should be invariant under the action of the AdS group
           $SO(2,2)$; \vsm
\item[(c)] asymptotic symmetries should have well defined canonical
           generators.
\eitem
The conditions (a) and (b) together lead to an extended asymptotic
structure, quite different from the standard, form-invariant vacuum
configuration, while (c) is just a technical assumption.

\prg{AdS asymptotics.} We begin our considerations with the point (a)
above. The asymptotic behaviour of the black hole triad \eq{3.2a} is
given by
$$
b^i{_\m}\sim \left( \ba{ccc}
        \dis \frac{r}{\ell}-\frac{m\ell}{r}  & 0 & 0      \\
        0 &\dis\frac{\ell}{r}+\frac{m\ell^3}{r^3}\,\, & 0 \\
        \dis\frac{J}{r}  & 0       & r
                    \ea
             \right)    \, ,
$$
where the type of higher order terms on the right hand side is not
written explicitly. Similarly, the asymptotic behaviour of the
connection \eq{3.2b} has the form
$$
\om^i{_\m}\sim \left( \ba{ccc}
    \dis\frac{p\ell}{2}\left(\frac{r}{\ell^2}-\frac{m}{r}\right) & 0
       &\dis -\frac{r}{\ell}+\frac{m\ell}{r}                       \\
     0 & \,\dis\frac{p\ell}{2r}+\frac{J\ell+pm\ell^3/2}{r^3}\, & 0 \\
    \dis-\frac{r}{\ell^2}+\frac{pJ}{2r} & 0
       & \dis \frac{pr}{2}-\frac{J}{r}
                      \ea
               \right) \, .
$$
According to (a), asymptotic conditions should be chosen so as to
{\it include\/} these black hole configurations.

In order to realize the requirement (b), we start with the above
black hole configuration and act on it with all possible $SO(2,2)$
transformations, defined by the basis of six pairs
($\xi_{(k)},\th_{(k)}$), displayed in Appendix A. The result has the
form
$$
\d_{(k)} b^i{_\m}\sim\left( \ba{ccc}
         \cO_1  & \cO_4 & \cO_1  \\
         \cO_2  & \cO_3 & \cO_2  \\
         \cO_1  & \cO_4 & \cO_1
                   \ea
                     \right) \, ,\qquad
\d_{(k)} \om^i{_\m}\sim \left( \ba{ccc}
         \cO_1  & \cO_4 & \cO_1  \\
         \cO_2  & \cO_3 & \cO_2  \\
         \cO_1  & \cO_4 & \cO_1
                   \ea
                        \right)\, ,
$$
where $\cO_n$ denotes a quantity that tends to zero as $1/r^n$ or
faster, when $r\to\infty$.

The family of the black hole triads obtained in this way is
parametrized by six real parameters, say $\s_i$; we denote it by
$\cB_6$. In order to have a set of asymptotic states which is
sufficiently large to {\it include\/} the whole $\cB_6$, we adopt the
following asymptotic form for the triad field:
\bsubeq\lab{4.1}
\be
b^i{_\m}= \left( \ba{ccc}
       \dis\frac{r}{\ell}+\cO_1   & O_4  & O_1  \\
       \cO_2 & \dis\frac{\ell}{r}+\cO_3  & O_2  \\
       \cO_1 & \cO_4                     & r+\cO_1
                 \ea
          \right)
\equiv \left( \ba{ccc}
       \dis\frac{r}{\ell} & 0    & 0  \\ [3pt]
       0 & \dis\frac{\ell}{r}    & 0  \\ [6pt]
       0 & 0                     & r
              \ea
       \right)+B^i{_\m}   \, .                             \lab{4.1a}
\ee
The real meaning of this expression and its relation to $\cB_6$ is
clarified by noting that any $c/r^n$ term in $\cB_6$ is transformed
into the corresponding $c(t,\vphi)/r^n$ term in \eq{4.1a}, i.e.
constants $c=c(\s_i)$ are promoted to functions $c(t,\vphi)$. Thus,
\eq{4.1a} is a natural generalization of $\cB_6$.

The triad family \eq{4.1a} generates the Brown--Henneaux asymptotic
form of the metric,
$$
g_{\m\n}=\left( \ba{ccc}
          \dis\frac{r^2}{\ell^2}+\cO_0  & \cO_3  & \cO_0 \\
          \cO_3 & \dis -\frac{\ell^2}{r^2}+\cO_4 & \cO_3 \\
          \cO_0 &  \cO_3  & -r^2+\cO_0
                \ea
         \right)
\equiv \left( \ba{ccc}
           \dis\frac{r^2}{\ell^2} & 0   & 0 \\ [3pt]
           0 & \dis -\frac{\ell^2}{r^2} & 0 \\ [6pt]
           0 & 0  & -r^2
              \ea
       \right)+G_{\m\n}\, ,
$$
but clearly, it is not uniquely determined by it (any Lorentz
transform of the triad produces the same metric).

Having found the triad asymptotics, we now use similar arguments to
find the needed asymptotic behavior for the connection:
\be
\om^i{_\m}=\left( \ba{ccc}
  \dis\frac{pr}{2\ell}+\cO_1 & \cO_2  &
                               \dis -\frac{r}{\ell}+\cO_1 \\
  \cO_2 & \dis\frac{p\ell}{2r}+\cO_3  & \cO_2             \\
  \dis-\frac{r}{\ell^2}+\cO_1 & \cO_2 & \dis \frac{pr}{2}+\cO_1
                  \ea
           \right)
\equiv  \left( \ba{ccc}
    \dis\frac{pr}{2\ell} & 0 &\dis -\frac{r}{\ell} \\ [3pt]
    0 & \dis\frac{p\ell}{2r}  & 0                  \\ [3pt]
    \dis-\frac{r}{\ell^2} & 0 & \dis \frac{pr}{2}
               \ea
        \right)+\Om^i{_\m} \, .                            \lab{4.1b}
\ee
\esubeq
Note that the choice $\om^0{}_1,\om^2{}_1=\cO_2$, adopted in
\eq{4.1b}, represents an acceptable generalization of the conditions
$\om^0{}_1,\om^2{}_1=\cO_4$, suggested by the form of
$\d_{(k)}\om^i{}_\m$ (compare also with the conditions \eq{C1}).

As we have seen, the requirements (a) and (b) are not sufficient for
a unique determination of the asymptotic behavior. Our choice of the
asymptotics was guided by the idea to obtain the {\it most general\/}
asymptotic behavior compatible with (a) and (b), with arbitrary
higher-order terms $B^i{_\m}$ and $\Om^i{_\m}$. Although $B^i{_\m}$
and $\Om^i{_\m}$ are arbitrary at this stage, certain relations among
them will be established latter (Appendix C), using some additional
requirements. One can verify that the asymptotic conditions \eq{4.1}
are indeed invariant under the action of the AdS group $SO(2,2)$. In
the next step, we shall examine whether there is any {\it higher\/}
symmetry structure in \eq{4.1}, which will be the real test of our
choice.

\prg{Asymptotic symmetries.} Having chosen the asymptotic conditions
in the form \eq{4.1}, we now wish to find the subset of gauge
transformations that respect these conditions. Acting on a specific
field satisfying \eq{4.1}, these transformations are allowed to
change the form of the non-leading terms $B^i{_\m}$, $\Om^i{_\m}$, as
they are arbitrary by assumption. Thus, the parameters of the
restricted gauge transformations are determined by the relations
\bsubeq\lab{4.2}
\bea
&&\ve^{ijk}\th_j b_{k\m}-(\pd_\m\xi^{\r})b^i{_\r}
        -\xi^\r\pd_\r b^i{}_\m=\d_0 B^i{_\m}\, ,           \lab{4.2a}\\
&&-\pd_\m\th^i+\ve^{ijk}\th_j\om_{k\m}-(\pd_\m\xi^\r)\om^i{_\r}
        -\xi^{\r}\pd_\r\om^i{}_{\m}=\d_0\Om^i{_\m}\, .     \lab{4.2b}
\eea
\esubeq
The transformations defined in this way differ from those that are
associated to the form-invariant vacuum configurations ($\d_0
b^i{_\m}=0$, $\d_0\om^i{_\m}=0$). The restricted gauge parameters are
determined as follows \cite{20}.

The symmetric part of \eq{4.2a} multiplied by $b_{i\n}$ (six
relations) yields the transformation rule of the metric:
$$
-(\pd_\m\xi^\r)g_{\n\r}-(\pd_\n\xi^\r)g_{\m\r}
-\xi^\r\pd_\r g_{\m\n}=\d_0 G_{\m\n} \, .
$$
Expanding $\xi^\m$ in powers of $r^{-1}$, we find the solution
of these equations as
\bsubeq\lab{4.3}
\bea
&&\xi^0=\ell\left[ T
  +\frac{1}{2}\left(\frac{\pd^2 T}{\pd t^2}\right)
              \frac{\ell^4}{r^2}\right] +\cO_4\, ,         \lab{4.3a}\\
&&\xi^2=S-\frac{1}{2}\left(\frac{\pd^2 S}{\pd\vphi^2}\right)
              \frac{\ell^2}{r^2}+\cO_4\, ,                 \lab{4.3b}\\
&&\xi^1=-\ell\left(\frac{\pd T}{\pd t}\right)r+\cO_1\, ,   \lab{4.3c}
\eea
where the functions $T(t,\vphi)$ and $S(t,\vphi)$ satisfy the
conditions
$$
\frac{\pd T}{\pd\vphi}=\ell\frac{\pd S}{\pd t}\, , \qquad
\frac{\pd S}{\pd\vphi}=\ell\frac{\pd T}{\pd t}\, .         \eqno(4.4)
$$
In \grl, these equations define the two-dimensional conformal group
at large distances \cite{3}.

The remaining three components of \eq{4.2a} determine $\th^i$:
\bea
&&\th^0=-\frac{\ell^2}{r}\pd_0\pd_2T+\cO_3\, ,             \nn\\
&&\th^2=\frac{\ell^3}{r}\pd_0^2T+\cO_3\, ,                 \nn\\
&&\th^1=\pd_2 T+\cO_2\, .                                  \lab{4.3d}
\eea
\esubeq\setcounter{equation}{4}

The conditions \eq{4.2b} produce no new limitations on the
parameters.

Introducing the light-cone coordinates $x^\pm=x^0/\ell\pm x^2$, the
conditions (4.4) can be written in the form
$$
\pd_\pm(T\mp S)=0 \, ,
$$
from which one easily finds the general solution for $T$ and $S$:
\be
T+S=g(x^+)\, ,\qquad  T-S=h(x^-)\, ,                       \lab{4.5}
\ee
where $g$ and $h$ are two arbitrary, periodic functions.

The commutator algebra of Poincar\'e gauge transformations \eq{2.1}
is closed: we have $[\d_0',\d_0'']=\d_0'''$, where
$\d'_0\equiv\d_0(\xi',\th')$ and so on, and the composition law
reads:
\bea
&&\xi'''{}^\m=\xi'{}^\r\pd_\r\xi''{}^\m
              -\xi''{}^\r\pd_\r\xi'{}^\m\, ,               \nn\\
&&\th'''{}^i=\ve^i{}_{mn}\th'{}^m\th''{}^n
             +\xi'\cdot\pd\th''{}^i-\xi''\cdot\pd\th'{}^i\,.\nn
\eea
Substituting here the restricted form of the parameters \eq{4.3} and
comparing the lowest order terms, we find the relations
\bea
&&T'''=T'\pd_2S''+S'\pd_2T''-T''\pd_2S'-S''\pd_2T'\, ,     \nn\\
&&S'''=S'\pd_2S''+T'\pd_2T''-S''\pd_2S'-T''\pd_2T'\, ,     \lab{4.6}
\eea
that are expected to be the composition law for $(T,S)$. To clarify
the situation, consider the restricted form of the gauge parameters
\eq{4.3}, and separate it into two pieces: the leading terms
containing $T$ and $S$, which define the $(T,S)$ transformations, and
the higher order terms that remain after imposing $T=S=0$, which
define the {\it residual\/} (or pure) gauge transformations. If the
relations \eq{4.6} are to represent the composition law for the
$(T,S)$ transformations, one has to check their consistency with
higher order terms in the commutator algebra. As one can verify, the
commutator of two $(T,S)$ transformations produces not only a $(T,S)$
transformation, with the composition law \eq{4.6}, but also an
additional, pure gauge transformation. However, pure gauge
transformations are irrelevant for our discussion of the conservation
laws. Indeed, as we shall see in section 6, they do not contribute to
the values of the conserved charges (their generators vanish weakly).
Thus, we are naturally led to correct the non-closure of the $(T,S)$
commutator algebra by introducing an improved definition of the {\it
asymptotic symmetry\/} \cite{3,26}:
\bitem
\item[\bul] the asymptotic symmetry group is defined as the factor group
of the gauge group determined by \eq{4.3}, with respect to the
residual gauge group.
\eitem
In other words, two asymptotic transformations are identified if they
have the same $(T,S)$ pairs, and any difference stemming from the
pure gauge terms is ignored. The asymptotic symmetry of our spacetime
coincides with the {\it conformal symmetry\/} (see section 6).

In conclusion, the set of asymptotic conditions \eq{4.1} is shown to
be invariant under the conformal symmetry group, which is much larger
then the original AdS group $SO(2,2)$. The resulting configuration
space respects the requirements (a) and (b) formulated at the
beginning of this section. The asymptotic structure of the whole
phase space, as well as the status of the last requirement (c), will
be examined in the next two sections.

\section{Gauge generator}
\setcounter{equation}{0}

In gauge theories, the presence of unphysical variables is closely
related to the existence of gauge symmetries. The best way to
understand the dynamical content of these symmetries is to explore
the related canonical generator, which acts on the basic dynamical
variables via the PB operation. To begin the analysis, we rewrite the
action \eq{2.4} as
\bea
&&I=\int d^3 x\ve^{\m\n\r}\left[ab^i{}_\m R_{i\n\r}
    -\frac{1}{3}\L\ve_{ijk}b^i{}_\m b^j{}_\n b^k{}_\r\right.\nn\\
&&\hspace{80pt} +\left.\a_3\left(\om^i{}_\m\pd_\n\om_i{}_\r
    +\frac{1}{3}\ve_{ijk}\om^i{}_\m\om^j{}_\n\om^k{}_\r\right)
    +\frac{1}{2}\a_4 b^i{}_\m T_{i\n\r}\right]\, .         \lab{5.1}
\eea

\prg{Hamiltonian and constraints.} The basic Lagrangian variables
$(b^i{}_\m,\om^i{}_\m)$ and the corresponding canonical momenta
$(\pi^i{}_\m,\Pi^i{}_\m)$ are related to each other through the set
of primary constraints:
\bea
&&\phi_i{}^0\equiv\pi_i{}^0\approx 0\, ,
       \hspace{87pt} \Phi_i{}^0\equiv\Pi_i{}^0\approx 0\, ,\nn\\
&&\phi_i{}^\a\equiv\pi_i{}^\a-\a_4\ve^{0\a\b}b_{i\b}\approx 0\, ,
\qquad\Phi_i{}^\a\equiv\Pi_i{}^\a-\ve^{0\a\b}\left(2a b_{i\b}
      +\a_3\om_{i\b}\right)\approx 0\, .                   \lab{5.2}
\eea
Explicit construction of the canonical Hamiltonian yields an
expression which is linear in unphysical variables, as expected:
\bea
&&\cH_c= b^i{}_0\cH_i+\om^i{}_0\cK_i+\pd_\a D^\a\, ,       \nn\\
&&\cH_i=-\ve^{0\a\b}\left(aR_{i\a\b}+\a_4T_{i\a\b}
        -\L\ve_{ijk}b^j{}_\a b^k{}_\b\right) \, ,          \nn\\
&&\cK_i=-\ve^{0\a\b}\left(aT_{i\a\b}+\a_3R_{i\a\b}
        +\a_4\ve_{ijk}b^j{}_\a b^k{}_\b\right) \, ,        \nn\\
&& D^\a=\ve^{0\a\b}\left[ \om^i{}_0\left( 2ab_{i\b}
        +\a_3\om_{i\b}\right)+\a_4b^i{}_0 b_{i\b}\right]\,.\nn
\eea
Going over to the total Hamiltonian,
\be
\cH_T=b^i{}_0\cH_i+\om^i{}_0\cK_i
   +u^i{}_\m\phi_i{}^\m+v^i{}_\m\Phi_i{}^\m+\pd_\a D^\a\, ,\lab{5.3}
\ee
we find that the consistency conditions of the sure primary
constraints $\pi_i{}^0$ and $\Pi_i{}^0$ yield the secondary
constraints:
\bsubeq\lab{5.4}
\be
\cH_i\approx0, \qquad \cK_i\approx0 \, .                   \lab{5.4a}
\ee
These constraints can be equivalently written in the form:
\be
T_{i\a\b}\approx p\ve_{ijk}b^j{}_\a b^k{}_\b, \qquad
R_{i\a\b}\approx q\ve_{ijk}b^j{}_\a b^k{}_\b\, .           \lab{5.4b}
\ee
\esubeq
The consistency of the remaining primary constraints $\phi^i{}_\a$
and $\Phi^i{}_\a$ leads to the determination of the multipliers
$u^i{}_\b$ and $v^i{}_\b$ (see Appendix B):
\bsubeq\lab{5.5}
\bea
&&u^i{}_\b+\ve^{ijk}\om_{j0}b_{k\b}-\nabla_\b b^i_0
  =p\ve^{ijk}b_{jo}b_{k\b}\, ,                             \nn\\
&&v^i{}_\b-\nabla_\b\om^i{}_0
  =q\ve^{ijk}b_{j0}b_{k\b}\, .                             \lab{5.5a}
\eea
Using the equations of motion $\dot b^i{}_\b=u^i{}_\b$ and
$\dot\om^i{}_\b=v^i{}_\b$, these relations reduce to the field
equations
\be
T^i{}_{0\b}\approx p\ve^{ijk}b_{j0}b_{k\b}\, ,\qquad
R^i{}_{0\b}\approx q\ve^{ijk}b_{j0}b_{k\b}\, .             \lab{5.5b}
\ee
\esubeq
The substitution of the determined multipliers \eq{5.5a} into \eq{5.3}
yields the final form of the total Hamiltonian:
\bsubeq
\bea
&&\cH_T=\hcH_T+\pd_\a\bD^\a\, ,                            \nn\\
&&\hcH_T=b^i{}_0\bcH_i+\om^i{}_0\bcK_i
         +u^i{}_0\pi_i{}^0+v^i{}_0\Pi_i{}^0\, ,            \lab{5.6a}
\eea
where
\bea
&&\bcH_i=\cH_i-\nabla_\b\phi_i{}^\b
  +\ve_{ijk}b^j{}_\b\left(p\phi^{k\b}+q\Phi^{k\b}\right)\,,\nn\\
&&\bcK_i=\cK_i-\nabla_\b\Phi_i{}^\b
              -\ve_{ijk}b^j{}_\b\phi^{k\b}\, ,             \nn\\
&&\bD^\a=D^\a+b^i{}_0\phi_i{}^\a+\om^i{}_0\Phi_i{}^\a\, .  \lab{5.6b}
\eea
\esubeq

Further investigation of the consistency procedure is facilitated by
observing that the secondary constraints $\bcH_i,\bcK_i$ obey the PB
relations \eq{B2}. One concludes that the consistency conditions of
the secondary constraints \eq{5.4} are identically satisfied, which
completes the Hamiltonian consistency procedure.

Complete classification of the constraints is given in the following
table.
\begin{center}
\doublerulesep 1.8pt
\begin{tabular}{lll}
\multicolumn{3}{l}{\hspace{16pt}Table 1. Classification of constraints}\\
                                                       \hline\hline
\rule{0pt}{12pt}
&~First class\phantom{x}&~Second class\phantom{x}\\
                                                       \hline
\rule[-1pt]{0pt}{15pt}
\phantom{x}Primary &~$\p_i{^0},\Pi_i{^0}$ &~$\phi_i{^\a},\Phi_i{}^{\a}$\\
                                                       \hline
\rule[-1pt]{0pt}{15pt}
\phantom{x}Secondary\phantom{x} &~$\bcH_i,\bcK_i$     &~             \\
                                                       \hline\hline
\end{tabular}
\end{center}
\medskip

\prg{Canonical gauge generator.} The results of the previous analysis
are sufficient for the construction of the gauge generator \cite{27}.
Starting from the primary first class constraints $\pi_i{}^0$ and
$\Pi_i{}^0$, one obtaines:
\bea
&&G[\eps]=\dot{\eps^i}\pi_i{}^0+\eps^i\left[\bcH_i
          -\ve_{ijk}\left(\om^j{}_0-pb^j{}_0\right)\pi^{k0}
          +q\ve_{ijk}b^j{}_0\Pi^{k0}\right] \, ,           \nn\\
&&G[\t]=\dot{\t^i}\Pi_i{}^0+\t^i\left[\bcK_i
        -\ve_{ijk}\left(b^j{}_0\pi^{k0}
        +\om^j{}_0\Pi^{k0}\right)\right]\, .               \lab{5.7}
\eea
The complete gauge generator is given by the expression
$G=G[\eps]+G[\t]$, and its action on the fields, defined by
$\d_0\phi=\{\phi,G\}$, has the form:
\bea
&&\d_0b^i{}_\m=\nabla_\m\eps^i-p\ve^i{}_{jk}b^j{}_\m\t^k
               +\ve^i_{jk}b^j{}_\m\t^k  \, ,               \nn\\
&&\d_0\om^i{}_\m=\nabla_\m\t^i-q\ve^i_{jk}b^j{}_\m\eps^k\,.\nn
\eea
This result looks more like a standard gauge transformation, with no
trace of the expected local Poincar\'e transformations. However,
after introducing the new parameters
$$
\eps^i=-\xi^\m b^i{}_\m\,,\qquad\t^i=-(\th^i+\xi^\m\om^i{}_\m)\, ,
$$
one easily obtains
\bea
&&\d_0b^i{}_\m=\d_\pgt b^i{}_\m
 -\xi^\r\left(T^i{}_{\m\r}-p\ve^{ijk}b_{j\m}b_{k\r}\right)\,,\nn\\
&&\d_0\om^i{}_\m=\d_\pgt\om^i{}_\m
 -\xi^\r\left(R^i{}_{\m\r}-q\ve^{ijk}b_{j\m}b_{k\r}\right)\,.\nn
\eea
Thus, {\it on-shell\/}, we have the transformation laws that are in
complete agreement with \eq{2.1}. Expressed in terms of the new
parameters, the gauge generator takes the form
\bea
&&G=-G_1-G_2 \, ,                                          \nn\\
&&G_1\equiv\dot\xi^\r\left(b^i{}_\r\pi_i{}^0
  +\om^i{}_\r\Pi_i{}^0\right)+\xi^\r\left[b^i{}_\r\bcH_i
  +\om^i{}_\r\bcK_i +(\pd_\r b^i_0)\pi_i{}^0
  +(\pd_\r\om^i{}_0)\Pi^i{}_0\right] \, ,                  \nn\\
&&G_2\equiv\dot\th^i\Pi_i{}^0+\th^i\left[\bcK_i
  -\ve_{ijk}\left(b^j{}_0\pi^{k0}
  +\om^j{}_0\Pi^{k0}\right)\right]\, ,                     \lab{5.8}
\eea
where the time derivatives $\dot b^i{}_\m$ and $\dot\omega^i{}_\m$
are shorts for $u^i{}_\m$ and $v^i{}_\m$, respectively. Note, in
particular, that the time translation generator is determined by the
total Hamiltonian:
$$
G\left[\xi^0\right]=-\dot\xi^0\left(b^i{}_0\pi_i{}^0
                    +\om^i{}_0\Pi_i{}^0\right)-\xi^0\hcH_T \, .
$$

In the above expressions, the integration symbol $\int d^3 x$ is
omitted for simplicity; later, when necessary, it will be restored.

\prg{Asymptotics of the phase space.} In order to extend the
asymptotic conditions \eq{4.1} to the canonical level, one should
determine an appropriate asymptotic behavior of the whole phase
space, including the momentum variables. This step is based on the
following general principle:
\bitem
\item[\bul] the expressions than vanish on shell should have an
arbitrary fast asymptotic decrease, as no solutions of the field
equations are thereby lost.
\eitem
By applying this principle to the primary constraints \eq{5.2}, one
finds the following asymptotic behavior of the momentum variables:
\bea
&&\pi_i{}^0=\hcO\, , \hspace{26pt}
  \pi_i{}^\a=\a_4\ve^{0\a\b}b_{i\b}+\hcO                   \nn\\
&&\Pi_i{}^0=\hcO\, , \qquad
  \Pi_i{}^\a=2a\ve^{0\a\b}b_{i\b}
  +\a_3\ve^{0\a\b}\om_{i\b}+\hcO\, .                       \lab{5.9}
\eea
We shall use this principle again in connection to the consistency
requiremets \eq{5.4b} and \eq{5.5b}, in order to {\it refine\/} the
general asymptotic canditions \eq{4.1} and \eq{5.9} (Appendix C).

\section{Canonical structure of the asymptotic symmetry}
\setcounter{equation}{0}

In this section, we study the influence of the adopted asymptotics on
the canonical structure of the theory: we construct the improved
gauge generators, examine their canonical algebra and prove the
conservation laws.

\prg{Improving the generators.} The canonical generator acts on
dynamical variables via the PB operation, which is defined in terms
of functional derivatives. A phase-space functional $F=\int d^2x
f(\phi,\pd\phi,\pi,\pd\pi)$ has well defined functional derivatives
if its variation can be written in the form
$
\d F=\int{\rm d}^2x\left[A(x)\d\phi(x)+B(x)\d\pi(x)\right]\, ,
$
where terms $\d\phi,_\m$ and $\d\pi,_\n$ are absent. In order to
ensure this property for our generator \eq{5.8}, we have to improve
its form by adding certain surface terms \cite{28}.

Let us start the procedure by examining the variations of $G_2$:
\bea
\d G_2&=&\th^i\d\bcK_i+R=\th^i\d\cK_i+\pd\hcO +R           \nn\\
&=&-2\ve^{0\a\b}\th^i\left(a\pd_\a\d b_{i\b}
   +\a_3\pd_\a\d\om_{i\b}\right)+\pd\hcO+R                 \nn\\
&=&-2\ve^{0\a\b}\pd_\a\left(a\th^i\d b_{i\b}
   +\a_3\th^i\d\om_{i\b}\right)+\pd\hcO+R=\pd\cO_2+R\, ,   \nn
\eea
where the last equality follows from the asymptotic relations
$\th^i\d b_{i\b},\th^i\d\om_{i\b}=\cO_2$. The total divergence term
$\pd\cO_2$ gives a vanishing contribution after integration, as
follows from the Stokes theorem:
$$
\int_{\cM_2} d^2x\pd_\a v^\a
=\int_{\pd\cM_2}v^\a df_\a=\int_0^{2\pi}v^1 d\vphi
 \qquad (df_\a=\ve_{\a\b}dx^\b) \, ,
$$
where the boundary of the spatial section $\cM_2$ of spacetime is
taken to be the circle at infinity, parametrized by the angular
coordinate $\vphi$. Thus, the boundary term for $G_2$ vanishes,
and $G_2$ is regular as it stands, without any correction.

Going over to $G_1$, we have:
\bea
\d G_1&=&\xi^\r\left(b^i{}_\r\d\bcH_i
   +\om^i{}_\r\d\bcK_i\right)+\pd\hcO + R                  \nn\\
&=&-2\ve^{0\a\b}\pd_\a\left[\xi^\r b^i{}_\r\d\left(a\om_{i\b}
   +\a_4b_{i\b}\right)+\xi^\r\om^i{}_\r\d\left(ab_{i\b}
   +\a_3\om_{i\b}\right)\right]+\pd\hcO+R \, .             \nn
\eea
Using the adopted asymptotic conditions, the preceding result leads
to
\bea
\d G_1&=&
  -\pd_\a\left(\xi^0\d\cE^\a+\xi^2\d\cM^\a\right)
  +\pd\cO_2+R                                              \nn\\
  &=&-\d\pd_\a\left(\xi^0\cE^\a+\xi^2\cM^\a\right)
              +\pd\cO_2 + R \, ,                           \nn
\eea
where
\bea
&&\cE^\a\equiv
   2\ve^{0\a\b}\left[\left(a+\frac{\a_3p}{2}\right)\om^0{}_\b
  +\left(\a_4+\frac{ap}{2}\right)b^0{}_\b+\frac{a}{\ell}b^2{}_\b
  +\frac{\a_3}{\ell}\om^2{}_\b\right]b^0{}_0\, ,           \nn\\
&&\cM^\a\equiv
  -2\ve^{0\a\b}\left[\left(a+\frac{\a_3p}{2}\right)\om^2{}_\b
  +\left(\a_4+\frac{ap}{2}\right)b^2{}_\b+\frac{a}{\ell}b^0{}_\b
  +\frac{\a_3}{\ell}\om^0{}_\b\right]b^2{}_2\, .           \lab{6.1}
\eea
Thus, the improved form of the complete gauge generator \eq{5.8}
reads:
\bea
&&\tG=G+\G\, ,                                             \nn\\
&&\G=-\oint df_\a\left(\xi^0\cE^\a+\xi^2\cM^\a\right)
    =-\int_0^{2\pi}d\vphi\left(\ell T\cE^1+S\cM^1\right)\,.\lab{6.2}
\eea
The adopted asymptotic conditions guarantee that $\tG$ is finite and
differentiable functional. The boundary term $\G$ depends only on $T$
and $S$, not on any pure gauge term in \eq{4.3}.

The improved time translation generator has the form
\bsubeq
\bea
&&\tG[\xi^0]=G[\xi^0]-E[\xi^0]\, ,                         \nn\\
&&E[\xi^0]\equiv\int_0^{2\pi}d\vphi\,\xi^0\cE^1\, .        \lab{6.3a}
\eea
For $\xi^0=1$, the generator $G$ reduces to $-\hH_T$, and the
corresponding boundary term has the meaning of energy:
\be
\tH_T=\hH_T+E \, ,\qquad
E=\int_0^{2\pi}d\vphi\,\cE^1 \, .                          \lab{6.3b}
\ee
\esubeq
The improved spatial rotation generator is given by
\bsubeq
\bea
&&\tG[\xi^2]=G[\xi^2]-M[\xi^2]\, ,                         \nn\\
&&M[\xi^2]\equiv\int_0^{2\pi}d\vphi\,\xi^2\cM^1\, ,        \lab{6.4a}
\eea
where $M$ is a finite integral. The boundary term for $\xi^2=1$,
\be
M=\int_0^{2\pi}d\vphi\,\cM^1 \, ,                          \lab{6.4b}
\ee
\esubeq
is the angular momentum of the system.

\prg{Canonical algebra.} The PB algebra of the improved generators
could be found by a direct calculation, but we shall rather use
another, more instructive method, based on the results of Refs.
\cite{29} and \cite{20}. Let us first recall that our improved
generator \eq{6.2} is a differentiable phase space functional that
preserves the asymptotic conditions \eq{4.1} and \eq{5.9}, hence, it
satisfies the conditions of the main theorem in Ref. \cite{29}.
Introducing a convenient notation, $\tG'\equiv\tG[T',S']$,
$\tG''\equiv\tG[T'',S'']$, the main theorem states that the PB
$\{\tG'',\tG'\}$ of two differentiable generators is itself a
differentiable generator. Taking into account that any differentiable
generator $\tG$ is defined only up to an additive, constant
phase-space functional $C$ (which does not change the action of $\tG$
on the phase space),  the main theorem leads directly to
\be
\left\{\tG'',\tG'\right\}= \tG'''+C''',                    \lab{6.5}
\ee
where the parameters of $\tG'''$ are defined by the composition law
\eq{4.6}, and $C'''$ is an unknown, field-independent functional,
$C'''\equiv C'''[T',S';T'',S'']$. The term $C'''$ is known as the
{\it central charge\/} of the PB algebra.

In order to calculate $C'''$, we note that
$\{\tG'',\tG'\}=\d_0'\tG''\approx\d_0'\G''$, where the weak equality
is a consequence of the fact that $\d'_0$ is a symmetry operation
that maps constraints into constraints. Combining this result with
$\tG'''\approx\G'''$,  Eq. \eq{6.5} implies
\bsubeq\lab{6.6}
\be
\d_0'\G''\approx \G'''+C''' \, .                           \lab{6.6a}
\ee
This relation determines the value of $C'''$ only weakly, but since
$C'''$ is a field-independent quantity, the weak equality is easily
converted into the strong one. The calculation of $\d_0'\G''$ is
based on the relations
\bea
&&\d_0\left(\ell\cE^1\right)=-\cM^1\pd_2T-\ell\cE^1\pd_2S
  -\pd_2\left(\cM^1T+\ell\cE^1S\right)                     \nn\\
&&\hspace{3.4cm}
  +(2a+\a_3p)\ell\pd_2^3S-2\a_3\pd_2^3T+\cO_2\, ,          \nn\\
&&\d_0\cM^1=-\ell\cE^1\pd_2T-\cM^1\pd_2S
  -\pd_2\left(\ell\cE^1T+\cM^1S\right)                     \nn\\
&&\hspace{3.4cm}
  +(2a+\a_3p)\ell\pd_2^3T-2\a_3\pd_2^3S+\cO_2\, ,          \nn
\eea
which follow from the refined asymptotic conditions derived in
Appendix C, and the transformation rules defined by the parameters
\eq{4.3}. Substituting the calculated expression for $\d'_0\G''$ into
\eq{6.6a} yields the following value for the central charge $C'''$:
\bea
C'''&=&(2a+\a_3p)\ell\int_0^{2\pi}d\vphi
      \left(\pd_2S''\pd_2^2T'-\pd_2S'\pd_2^2T''\right)     \nn\\
&&-2\a_3\int_0^{2\pi}d\vphi
      \left(\pd_2T''\pd_2^2T'+\pd_2S''\pd_2^2S'\right)\, . \lab{6.6b}
\eea
\esubeq

\prg{Conservation laws.} As we noted in section 5, the improved total
Hamiltonian is one of the generators, $\tG[1,0]=-\ell\tH_T$. A direct
calculation based on the PB algebra \eq{6.5} shows that the
asymptotic generator $\tG[T,S]$ is conserved \cite{20}:
\be
\frac{d}{dt}\tG=\frac{\pd}{\pd t}\tG+\left\{\tG,\,\tH_T\right\}
  \approx \frac{\pd}{\pd t}\G[T,S]
         -\frac{1}{\ell}\G[\pd_2 S,\pd_2 T]= 0\, .         \lab{6.7}
\ee
This also implies the conservation of the boundary term $\G$.

Now, we wish to clarify the meaning of the conserved charges by
calculating their values for the black hole solution \eq{3.2}. First,
note that the black hole solution depends on the radial coordinate
only, and consequently, the terms $\cE^1$ and $\cM^1$ in $\G$ behave
as constants. Second, the parameters ($T,S$) are periodic functions,
equation \eq{4.5}, so that only {\it zero modes\/} in the Fourier
expansion of ($T,S$) survive the integration in $\G$. Thus, there are
only two independent non-vanishing charges for the black hole
solution, given by two inequivalent choices of the {\it constants\/}
$T$ and $S$. If we take, for instance, ($T=1,S=0$) as the first
choice, and ($T=0,S=1$) as the second one, all the other non-zero
charges will be given as linear combinations of these two.

For ($T=1,S=0$) we have $\G[1,0]=-\ell E$, and the corresponding
conserved charge is the energy $E$. Its value for the black hole
solution is found to be
\bsubeq\lab{6.8}
\be
E(\mbox{black hole})=4\pi\left[m\left(a+\frac{\a_3p}{2}\right)
      -\frac{\a_3J}{\ell^2}\right]\, .                     \lab{6.8a}
\ee
The second choice ($T=0,S=1$) leads to $\G[0,1]=-M$. The
corresponding conserved charge is the angular momentum $M$, and its
black hole value reads
\be
M(\mbox{black hole})=4\pi\left[J\left(a+\frac{\a_3p}{2}\right)
      -\a_3m\right]\, .                                    \lab{6.8b}
\ee
\esubeq

Our expressions for the conserved charges \eq{6.8} coincide with the
results obtained in Ref. \cite{19}. In the sector $\a_3=0$ (\grl\ and
the teleparallel theory), we have $E=m$ and $M=J$ (in units $4G=1$),
while for $\a_3\ne 0$, the constants $m$ and $J$ do not have directly
the meaning of energy and angular momentum, respectively.
Geometrically, the two independent charges \eq{6.8} parametrize the
family of globally inequivalent, asymptotically AdS spaces.

\prg{Central charge.} Using the Fourier expansion, one can rewrite
the canonical algebra \eq{6.5} in a more familiar form.  The
parameters ($T,S$) can be Fourier decomposed as follows:
$$
T=\sum_{-\infty}^{+\infty}
  \left(a_ne^{inx^+}+\bar{a}_ne^{inx^-}\right)\, ,\qquad
S=\sum_{-\infty}^{+\infty}
  \left(a_ne^{inx^+}-\bar{a}_ne^{inx^-}\right)\, .
$$
The asymptotic generator is a linear, homogeneous function of the
parameters, so that:
$$
\tG[T,S]=-2\sum_{-\infty}^{+\infty}
         \left(a_nL_n+\bar{a}_n\bar{L}_n\right)\, .
$$
The previous relations imply:
\bsubeq
\be
2L_n=-\tG[T=S=e^{inx^+}]\, ,\qquad
2\bL_n=-\tG[T=-S=e^{inx^-}].                               \lab{6.9a}
\ee
Expressed in terms of the Fourier coefficients $L_n$ and $\bar L_n$,
the canonical algebra takes the form of two independent Virasoro
algebras with classical central charges:
\bea
&&\left\{L_n,L_m\right\}=-i(n-m)L_{n+m}
                 -\frac{c}{12}in^3\d_{n,-m}\, ,            \nn\\
&&\left\{\bL_n,\bL_m\right\}=-i(n-m)\bL_{m+n}
                  -\frac{\bar c}{12}in^3\d_{n,-m}\, ,      \nn\\
&&\{L_n,\bar{L}_m\}=0\, .                                  \lab{6.9b}
\eea
\esubeq
The central charges, given in the standard string theory normalization,
have the form:
\bea
&&c=12\cdot2\pi
   \left[a\ell+\a_3\left(\frac{p\ell}{2}-1\right)\right]\, ,\nn\\
&&\bar{c}=12\cdot2\pi
   \left[a\ell+\a_3\left(\frac{p\ell}{2}+1\right)\right]\,.\lab{6.10}
\eea
Thus, the gravitational sector with $\a_3\ne 0$ has the conformal
asymptotic symmetry with two different central charges, while
$\a_3=0$ implies $c=\bar c=3\ell/2G$.
\bitem
\item[\bul] The general classical central charges $c$ and $\bar c$
differ from each other, in contrast to the results obtained in \grl\
and the teleparallel theory \cite{3,20}.
\eitem

By redefining the zero modes, $L_0\ra L_0+c/24$, $\bL_0\ra \bL_0+\bar
c/24$, the Virasoro algebra takes its standard form. One should note
that the central term for the  $SO(2,2)$ subgroup, generated by
$(L_{-1},L_0,L_1)$ and $(\bL_{-1},\bL_0,\bL_1)$, vanishes. This is a
consequence of the fact that $SO(2,2)$ is an exact symmetry of the AdS
vacuum \cite{3}.

\section{Concluding remarks}
In this paper, we investigated the canonical structure of 3D gravity
with torsion.

(1) The geometric arena for the topological 3D gravity with torsion,
defined by the Mielke-Baekler action \eq{2.4}, has the form of
Riemann-Cartan spacetime.

(2) There exists an exact vacuum solution of the theory, the
Riemann-Cartan black hole \eq{3.2}, which generalizes the standard
BTZ black hole in \grl.

(3) Assuming the AdS asymptotic conditions, we constructed the
canonical conserved charges. Energy and angular momentum of the
Riemann-Cartan black hole are different from the corresponding BTZ
values.

(4) The PB algebra of the canonical generators has the form of two
independent Virasoro algebras with classical central charges. The
values of the central charges are different from each other, in
contrast to the situation in \grl\ and the teleparallel theory. The
implications of this result for the quantum structure of black hole
are to be explored.

\section*{Acknowledgements}

This work was supported by the Serbian Science foundation, Serbia.
One of us (MB) would like to thank Milovan Vasili\'c, Friedrich  Hehl
and Yuri Obukhov for a critical reading of the manuscript and many
useful suggestions.

\appendix
\section{Symmetries of the AdS vacuum}
\setcounter{equation}{0}

The invariance conditions $\d_0 b^i{}_\m=0$ for the AdS triad
\eq{3.3a} yield the set of requirements on the parameters
$(\xi^\m,\th^i)$, the general solution of which has the form
\cite{20}
\bea
&&\xi^0=\ell \s_1-\frac{r}{f}\pd_2Q \, ,\qquad
  \xi^1= \ell^2 f\pd_0\pd_2Q\, ,\qquad
  \xi^2=\s_2-\frac{\ell^2 f}{r}\pd_0Q \,,                  \nn\\
&&\th^0=-\frac{\ell^2}{r}\,\pd_0Q\, ,\hspace{42pt}
  \th^1=Q\, ,\hspace{62pt}
  \th^2=\frac{1}{f}\,\pd_2Q\,,                             \lab{A1}
\eea
where
\be
Q\equiv \s_3\cos x^+ +\s_4\sin x^+
       +\s_5\cos x^- +\s_6\sin x^-\, ,                     \lab{A2}
\ee
and $\s_i$ are six arbitrary dimensionless parameters. The invariance
conditions $\d_0\om^i{}_\m=0$ for the AdS connection \eq{3.3b} do not
produce any new restrictions on ($\xi^\m,\th^i$). For each
$k=1,2,\dots,6$, we can choose $\s_k=1$ as the only non-vanishing
constant, and find the corresponding basis of six independent Killing
vectors $\xi^\m_{(k)}$:
\bea
&&\xi_{(1)}=(\ell,0,0)\, ,                                 \nn\\
&&\xi_{(2)}=(0,0,1)\, ,                                    \nn\\
&&\xi_{(3)}=\left(\frac{r}{f}\,\sin x^+,
  -\ell f\cos x^+,\frac{\ell f}{r}\sin x^+ \right)\, ,     \nn\\
&&\xi_{(4)}=\left(\frac{r}{f}\,\cos x^+,
   \ell f\sin x^+,\frac{\ell f}{r}\cos x^+ \right)\, ,     \nn\\
&&\xi_{(5)}=\left(-\frac{r}{f}\,\sin x^-,
   \ell f\cos x^-,\frac{\ell f}{r}\sin x^- \right)\, ,     \nn\\
&&\xi_{(6)}=\left(\frac{r}{f}\,\cos x^-,
   \ell f\sin x^-,-\frac{\ell f}{r}\cos x^- \right)\, ,    \lab{A3}
\eea
and similarly for $\th^i_{(k)}$. As one can explicitly verify, the
six pairs ($\xi^\m_{(k)},\th^i_{(k)}$) fall into the class of
asymptotic parameters (4.3), and define the algebra of the AdS group
$SO(2,2)$.

\section{The algebra of constraints}
\setcounter{equation}{0}

The structure of the PB algebra of constraints is an important
ingredient in the  analysis of the Hamiltonian consistency
conditions. For the nontrivial part of the PB algebra involving
($\phi_i{}^\a,\Phi_i{}^\a,\cH_i,\cK_i$), we have the following
result:
\bea
&&\{\phi_i{}^\a,\phi_j{}^\b\}
  = -2\a_4\ve^{0\a\b}\eta_{ij}\d\, ,  \qquad
  \{\phi_i{}^\a,\Phi_j{}^\b\}=-2a\ve^{0\a\b}\eta_{ij}\d\, ,\nn\\
&&\{\Phi_i{}^\a,\Phi_j{}^\b\}
  = -2\a_3\ve^{0\a\b}\eta_{ij}\d\, ,                       \nn\\
&&\{\phi_i{}^\a,\cH_j\}
 =2\ve^{0\a\b}\left[\a_4\eta_{ij}\pd_\b\d
  -\ve_{ijk}\left(\a_4\om^k{}_\b-\L b^k{}_\b\right)\d\right]\,,\nn\\
&&\{\phi_i{}^\a,\cK_j\}
 =2\ve^{0\a\b}\left[a\eta_{ij}\pd_\b\d
  -\ve_{ijk}\left(a\om^k{}_\b+\a_4b^k{}_\b\right)\d\right]\,,\nn\\
&&\{\Phi_i{}^\a,\cH_j\}
 =2\ve^{0\a\b}\left[a\eta_{ij}\pd_\b\d
  -\ve_{ijk}\left(a\om^k{}_\b+\a_4b^k{}_\b\right)\d\right]\,,\nn\\
&&\{\Phi_i{}^\a,\cK_j\}
 =2\ve^{0\a\b}\left[\a_3\eta_{ij}\pd_\b\d
  -\ve_{ijk}\left(\a_3\om^k{}_\b+ab^k{}_\b\right)\d\right]\,.\lab{B1}
\eea
The essential part of the PB algebra involving the first class
constraints ($\bcH_i,\bcK_i$) is given by the following relations:
\bea
&&\{\phi_i{}^\a,\bcH_j\}
  =\ve_{ijk}\left(p\phi^{k\a}+q\Phi^{k\a}\right)\d\,,\qquad
   \{\phi_i{}^\a,\bcK_j\}=-\ve_{ijk}\phi^{k\a}\d\, ,       \nn\\
&&\{\Phi_i{}^\a,\bcH_j\}=-\ve_{ijk}\phi^{k\a}\d\, ,
\hspace{75pt}\{\Phi_i{}^\a,\bcK_j\}=-\ve_{ijk}\Phi^{k\a}\d\,,\nn\\
&&\{\bcH_i,\bcH_j\}=\ve_{ijk}\left(p\bcH^k+q\bcK^k\right)\d\, ,
\hspace{35pt} \{\bcH_i,\bcK_j\}=-\ve_{ijk}\bcH^k\d\, ,     \nn\\
&&\{\bcK_i,\bcK_j\}=-\ve_{ijk}\bcK^k\d\, .                 \lab{B2}
\eea

\section{Asymptotic form of the constraints}
\setcounter{equation}{0}

Here, we analyze the influence of the secondary constraints \eq{5.4}
and relations \eq{5.5} for the determined multipliers, on the basic
asymptotic conditions \eq{4.1} and \eq{5.9}, using the principle
formulated at the end of section 5.

Let us start with the secondary constraints \eq{5.4b}. Using \eq{4.1}
and \eq{5.9}, these constraints imply the following asymptotic
relations:
\bea
&& \om^0{}_1=\cO_4\, ,\hspace{41pt} \om^2{}_1=\cO_4\, ,    \nn\\
&&\pd_1(re_2)=\cO_3,\qquad \pd_1(rm_2)=\cO_3\, ,           \nn\\
&&\pd_1\left[r(B^2{}_2-B^0{}_2)\right]
  =\ell\left(\Om^2{}_2-\Om^0{}_2\right)+r^2\Om^1{}_1       \nn\\
&&\hspace{100pt}
   +\left(1-\frac{p\ell}{2}\right)
   \left(B^2{}_2-B^0{}_2+\frac{r^2}{\ell}B^1{}_1\right)
   +\cO_3 \, ,                                             \nn\\
&&\pd_1\left(rB^2{}_2\right)=\frac{p\ell}2B^0{}_2+B^2{}_2
  +\frac{r^2}{\ell}B^1{}_1-\ell\Om^0{}_2+\cO_3\, ,         \lab{C1}
\eea
where:
\bea
&&e_\m=\left(a+\frac{\a_3p}{2}\right)\om^0{}_\m
 +\left(\a_4+\frac{ap}{2}\right)b^0{}_\m
 +\frac{a}{\ell}b^2{}_\m+\frac{\a_3}{\ell}\om^2{}_\m\, ,   \nn\\
&&m_\m=\left(a+\frac{\a_3p}{2}\right)\om^2{}_\m
 +\left(\a_4+\frac{ap}{2}\right)b^2{}_\m
 +\frac{a}{\ell}b^0{}_\m+\frac{\a_3}{\ell}\om^0{}_\m\, .   \nn
\eea
From the expressions \eq{C1}, one easily concludes that the terms
$\cE^\a$ and $\cM^\a$, included in the surface integral \eq{6.2} for
$\G$, satisfy the following asymptotic conditions:
\be
\pd_1\cE^1=\cO_3\, , \qquad \cE^2=\cO_3\, ,\qquad
\pd_1\cM^1=\cO_3, \qquad \cM^2=\cO_3 \, .                  \lab{C2}
\ee

In a similar manner, equations \eq{5.5b} lead to:
\bea
&&\pd_1\left(rB^0{}_0\right)=B^0{}_0+\frac{p\ell}2B^2{}_0
  +\frac{r^2}{\ell^2}B^1{}_1-\ell\Om^2{}_0+\cO_3\, ,       \nn\\
&&\pd_1(re_0)=\cO_3\, ,\hspace{44pt} \pd_1(rm_0)=\cO_3\, , \nn\\
&&\pd_2 e_0-\pd_0 e_2=\cO_3\, ,\qquad
  \pd_2 m_0-\pd_0 m_2=\cO_3\, ,                            \nn\\
&&\ell e_0+m_2=\cO_3\, ,\hspace{35pt}\ell m_0+e_2=\cO_3\, .\lab{C3}
\eea

\end{document}